\documentclass[twocolumn,english,nofootinbib,superscriptaddress,pra,preprintnumbers,amsmath,amssymb,floatfix,longbibliography]{revtex4-1}
\usepackage[T1]{fontenc}
\usepackage[utf8]{inputenc}
\usepackage{xcolor}
\usepackage{babel}
\usepackage{bm}
\usepackage{amsmath}
\usepackage{graphicx}
\usepackage{esint}
\usepackage[unicode=true,pdfusetitle,
 bookmarks=true,bookmarksnumbered=false,bookmarksopen=false,
 breaklinks=false,pdfborder={0 0 0},pdfborderstyle={},backref=false,colorlinks=true]
 {hyperref}

\makeatletter
\newcommand{\KSadded}[1]{\textcolor{blue}{#1}}

\usepackage{dcolumn}
\usepackage{bm}
\usepackage{color}
\usepackage{soul}
\usepackage{babel}
\usepackage{amsfonts}
\usepackage{slashed}
\usepackage{enumerate}

\usepackage{babel}

\makeatother

\begin{document}
\global\long\def\sgn{\mathrm{sgn}}%
\global\long\def\ket#1{\left|#1\right\rangle }%
\global\long\def\bra#1{\left\langle #1\right|}%
\global\long\def\sp#1#2{\langle#1|#2\rangle}%
\global\long\def\abs#1{\left|#1\right|}%
\global\long\def\avg#1{\langle#1\rangle}%

\title{Quantum Zeno effect with partial measurement and noisy dynamics}
\author{Parveen Kumar}
\affiliation{Department of Condensed Matter Physics, Weizmann Institute of Science,
Rehovot 76100, Israel}
\author{Alessandro Romito}
\affiliation{Department of Physics, Lancaster University, Lancaster LA1 4YB, United
Kingdom}
\author{Kyrylo Snizhko}
\affiliation{Department of Condensed Matter Physics, Weizmann Institute of Science,
Rehovot 76100, Israel}
\begin{abstract}
We study the Quantum Zeno Effect (QZE) induced by continuous partial
measurement in the presence of short-correlated noise in the system
Hamiltonian. We study the survival probability and the onset of the
QZE as a function of the measurement strength, and find that, depending
on the noise parameters, the quantum Zeno effect can be enhanced or
suppressed by the noise in different regions of the parameter space.
Notably, the conditions for the enhancement of the QZE are different
when determined by the short-time or long-time behavior of the survival
probability, or by the measurement strength marking the onset of the
quantum Zeno regime.
\end{abstract}
\maketitle

\section{Introduction}

The evolution of a quantum system can be ``frozen'' by repeated
projective measurements; this phenomenon is known as quantum Zeno
effect (QZE) \citep{Misra1977,Itanoa,Facchi2008}. The QZE originates
from the general feature of Schr\"odinger evolution: at short times,
transition probabilities between quantum states are quadratic in time;
hence if a system in a measurement eigenstate is measured at regular
intervals $\Delta t$, the probability of a transition to another
state is asymptotically small in the limit $\Delta t\to0$ \citep{Home1997,NakazatoHiromichiNamikiMikioandPascazio1996}.
The QZE is a well established experimental reality: it has been explored
experimentally in various setups such as trapped ions \citep{Itanoa},
polarized photons \citep{Kwiat1995}, cold atoms \citep{Fischer2001},
dilute Bose--Einstein condensed gases \citep{Streed2006}, nanomechanical
oscillators \citep{Chen2010}, and superconducting qubits \citep{Slichter2016}.
Besides confining a system in a specific state, the QZE can be used
for a few other purposes such as stabilizing a multidimensional subspace
\citep{Facchi2002,Facchi2008} (a feature which has been experimentally
observed with a rubidium Bose--Einstein condensate \citep{Schafer2014,Signoles2014}
and that can be useful for quantum error correction \citep{Beige2000,Maniscalco2008,Kim2012,Paz-Silva2012,Chen2020})
or tuning the degree of Markovianity in the dynamics of an open quantum
system \citep{Patsch2020}. Further, QZE can play an important role
in the dynamics of many-body systems \citep{Tonielli2020a,Dolgirev2020}.

Given the potential applications of the QZE, it is important to study
the effect under realistic non-ideal conditions, in which uncertainties
in the system Hamiltonian and in the measurement process are unavoidable.
For example, recent studies addressed the QZE due to randomly-spaced
projective measurements \citep{Shushin2011,Gherardini2016} or under
non-projective (\emph{generalized}) measurements \citep{Gong2004,Koshino2005,Streed2006,Xiao2006,Ruskov2006,Breuer2007,Paz-Silva2012,Layden2015,Zhang2019}.
Interestingly, it has been found that a complete stabilization of
the system in specific states is possible even with imperfect measurements
occurring at finite frequencies \citep{Layden2015}.

\begin{figure}
\begin{centering}
\includegraphics[width=1\columnwidth]{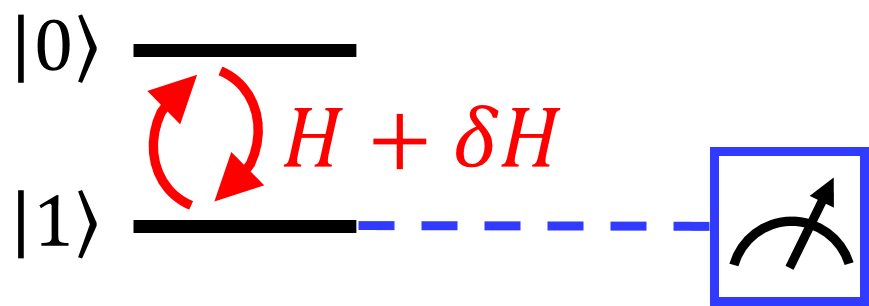}
\par\end{centering}
\caption{\label{fig:setup} The setup under consideration. A two-level system
is subject to a fluctuating Hamiltonian that induces quantum oscillations
between the two levels. Simultaneously, the system is subject to a
continuous partial measurement: the detector has a finite probability
per unit time to click when the system is in state $\protect\ket 1$,
making the detection of the system state imperfect.}
\end{figure}

In this work, we study a different kind of non-ideality. While the
above-mentioned works assumed a fully known and controlled Hamiltonian,
we consider a system with a noisy Hamiltonian subjected to continuous
partial measurements \citep{Elitzur2001,Paraoanu2006,Ruskov2007,Xu2011,Blok2014},
cf.~Fig.~\ref{fig:setup}. Several specific types of noise have
been previously considered \citep{Nakazato1999,Kofman1999,Kofman2001,Gurvitz2003}.
In particular, it has been shown that in the presence of short-correlated
noise the system dynamics cannot be frozen completely \citep{Gurvitz2003}.
Nevertheless, the presence of measurement can slow down the system's
departure from the desired state. We investigate whether the QZE (in
the sense of slowing the system decay down) is enhanced or suppressed
by an \emph{arbitrary} short-correlated noise in the Hamiltonian as
compared to the case in which the noise is absent. We use three distinct
observables to address the effect of noise in this regard: the short-time
and the long-time survival probabilities, as well as the critical
value of the measurement strength for the onset of the QZE. The latter
is identified via a transition from the regime in which population
of the state of interest oscillates in time to the regime in which
it monotonically decays (with the decay rate vanishing for perfect
QZE) \citep{Li2014a,Snizhko2020d}. We find that the noise can enhance
the QZE. At the same time, the conditions for enhancing the QZE behavior
differ drastically for each of the observables. While the short-time
probability of staying (``surviving'') in the initial state is essentially
always reduced by noise, the long-time survival probability is \emph{increased}
in some parameter range, specifically when the measurement strength
and the average Hamiltonian parameters fall within a certain noise-determined
region. Finally, we identify a set of constraints for the noise strength
and average Hamiltonian parameters which extend the QZE regime, by
inducing the transition to the QZE at a \emph{smaller} measurement
strength. Notably, these constraints, do not exactly match the ones
determining the region of the long-time survival probability enhancement.

This paper is organized as follows. Section~\ref{sec:The-setup}
introduces the setup under consideration and its modeling. The features
of the QZE in our setup when the Hamiltonian is noiseless are reviewed
in Section~\ref{sec:in_the_absence_of_any_noise}. In Sec.~\ref{sec:the_effect_of_noise},
we investigate the effect of noise on the QZE. We conclude with a
discussion of our results in Section~\ref{sec:conclusion}.

\section{The setup\label{sec:The-setup}}

We consider a two-level system, whose basis states are labeled as
$\ket 0$ and $\ket 1$, cf.~Fig.~\ref{fig:setup}. We represent
the system state by the density matrix

\begin{equation}
\rho_{s}(t)=\frac{1}{2}(\mathbb{I}+\bm{s}\cdot\bm{\sigma}),\label{eq:system_gen_state}
\end{equation}
where $\bm{s}=\{x(t),y(t),z(t)\}$ is the Bloch vector, $\mathbb{I}$
is the identity operator, and $\bm{\sigma}=\{\sigma_{x},\sigma_{y},\sigma_{z}\}$
are the Pauli spin operators defined by $\sigma_{z}\ket 0=\ket 0$.
We will consider the system's initial state to be
\begin{equation}
\rho_{s}(0)=\ket 0\bra 0=\begin{pmatrix}1 & 0\\
0 & 0
\end{pmatrix}.\label{eq:system_initial_state}
\end{equation}
The system evolution is dictated by the interplay of two ingredients:
a Hamiltonian including an explicit noise component and the back-action
of the measurements. We describe their effect separately, in Secs.~\ref{sec:setup_Hamiltonian+noise}
and \ref{sec:setup_measurement-model}, before combining them to derive
the full master equation for the system's density matrix in Sec.~\ref{sec:setup_combined-evolution}.

\subsection{\label{sec:setup_Hamiltonian+noise}System Hamiltonian and noise}

In the absence of measurements, the system undergoes a unitary evolution
with the Hamiltonian

\begin{equation}
H(t)=\omega\,\sigma_{x}+\bm{\xi}(t)\cdot\bm{\sigma},\label{eq:total_Hamiltonian}
\end{equation}
where $\omega$ drives Rabi oscillations between the two levels, and
$\bm{\xi}(t)$ is a time-dependent short-correlated noise with a Gaussian
distribution such that $\langle\xi_{i}(t)\rangle=0$ and $\langle\xi_{i}(t)\xi_{j}(t')\rangle=\gamma_{ij}\delta(t-t')$,
where $i\in\left\{ x,y,z\right\} $. Here $\gamma_{ij}$ is a real
positive semi-definite matrix and $\delta(t)$ is the Dirac delta
function. In what follows, we discretize the evolution into time steps
of size $dt$, in which case $\delta(t-t')=\delta_{t,t'}/dt$, where
$\delta_{t,t'}$ is the Kronecker delta. Each individual realization
of the noise generates a stochastic trajectory of the system state
evolution on the Bloch sphere and the system state remains pure during
the whole evolution ($\bm{s}^{2}(t)=1$). After the averaging over
different noise realizations, the system state becomes mixed, and
the corresponding Bloch vector points to the interior of the Bloch
sphere ($\bm{s}^{2}(t)<1$). In this work, we will be interested in
the averaged evolution of the system state.

The system state evolution for one time step, averaged over the noise,
is given by

\begin{multline}
\rho_{s}(t+dt)=\left\langle U_{t}\rho_{s}(t)U_{t}^{\dagger}\right\rangle _{\bm{\xi}(t)}=\rho_{s}(t)-i\omega\left[\sigma_{x},\rho_{s}(t)\right]dt\\
-\gamma_{ij}\left(\frac{1}{2}\left\{ \sigma_{i}\sigma_{j},\rho_{s}(t)\right\} -\sigma_{i}\rho_{s}(t)\sigma_{j}\right)dt+O(dt^{2}),\label{eq:system_evol_with_sys_Ham}
\end{multline}
where $U_{t}=e^{-iH(t)dt}$, $\left[..,..\right]$ and $\left\{ ..,..\right\} $
stand for the commutator and the anti-commutator respectively. 

\subsection{\label{sec:setup_measurement-model}The measurement model}

The measurement consists of continuous partial measurements, cf.~Fig.
\ref{fig:setup}. Depending on the probability of detecting the system
in $\ket 1$ state, the measurements can bridge between no measurement
(vanishing back-action) and the projective measurement. Continuous
partial measurements have been realized in various experimental architectures
\citep{Kim2012,Minev2019}.

A simple physical model of partial measurements is obtained by considering
a two-level-system detector (with basis states $\ket 0_{d}$, $\ket 1_{d}$)
initialized at the beginning of each measurement in state $|\psi\rangle_{d}=|0\rangle_{d}$.
The joint system-detector state before the measurement can thus be
written as

\begin{equation}
\rho(t)=\rho_{s}(t)\otimes\left(|\psi\rangle_{d}\langle\psi|_{d}\right).\label{eq:sys_det_joint_state}
\end{equation}
During the measurement, the system-detector interaction,

\begin{equation}
H_{int}=\frac{J}{2}(\mathbb{I}-\sigma_{z})\otimes\sigma_{y}^{(d)},\label{eq:sys_det_int_Hamiltonian}
\end{equation}
is switched on for time interval $dt$; here $J$ determines the coupling
strength between the system and the detector. The entangled system-detector
state after the time step $dt$ is given by

\begin{equation}
\rho(t+dt)=V\rho(t)V^{\dagger},\label{eq:meas_unitary_evolution}
\end{equation}
where $V=e^{-iH_{int}dt}$. At the end, the detector state is measured
projectively in the $\{|0\rangle_{d},|1\rangle_{d}\}$ basis. The
measurement procedure is then repeated, giving rise to continuous
partial measurement. 

The system state after the measurement is obtained by tracing over
the detector degrees of freedom,

\begin{multline}
\rho_{s}(t+dt)=\textrm{Tr}_{d}[\rho(t+dt)]\\
=M_{0}\rho_{s}(t)M_{0}^{\dagger}+M_{1}\rho_{s}(t)M_{1}^{\dagger},\label{eq:post_meas_system_state}
\end{multline}
where $M_{r}$ are the Kraus operators \citep{Nielsen2010,Wiseman2010}
encoding the measurement back-action corresponding to a particular
readout $r=0,1$, i.e., when the detector is found to be in state
$\ket r_{d}$ at the end of the measurement. They are given by \KSadded{$M_r={}_d \langle r|V|0\rangle_d$}.
Using Eq.~(\ref{eq:sys_det_int_Hamiltonian}), the Kraus operators
$M_{0}$ and $M_{1}$ can be written as
\begin{equation}
M_{0}=\begin{pmatrix}1 & 0\\
0 & \cos(Jdt)
\end{pmatrix},\quad\ensuremath{M_{1}=}\begin{pmatrix}0 & 0\\
0 & \sin(Jdt)
\end{pmatrix}.\label{eq:gen_meas_op_M0_and_M1}
\end{equation}

\KSadded{Equations~(\ref{eq:post_meas_system_state}) and (\ref{eq:gen_meas_op_M0_and_M1})}
are the defining properties of the partial measurement used hereafter.
They are independent of the specific detector model and could equivalently
result from a measurement process different from the one used to illustrate
their derivation.

Note that the measurement strength is quantified by $Jdt$ such that
$Jdt=0$ corresponds to no measurement, while $Jdt=\pi/2$ corresponds
to strong (or projective) measurement. We obtain the continuum limit,
$dt\rightarrow0$, by scaling $J$ such that $J^{2}dt=\mathrm{const}\equiv\alpha$.
The master equation giving the evolution of the system density matrix
under continuous partial measurement then follows as
\begin{multline}
\rho_{s}(t+dt)=\rho_{s}(t)-\alpha\left(\frac{1}{2}\left\{ P_{1},\rho_{s}(t)\right\} -P_{1}\rho_{s}(t)P_{1}\right)dt\\
+O(dt^{2}),
\end{multline}
where $P_{1}=\ket 1\bra 1=(\mathbb{I}-\sigma_{z})/2$.

\subsection{\label{sec:setup_combined-evolution}Combined system evolution}

The situation of interest for us is that the Hamiltonian, cf.~Sec.~\ref{sec:setup_Hamiltonian+noise},
and the measurement-induced, cf.~Sec.~\ref{sec:setup_measurement-model},
evolutions happen simultaneously. In a single small time step, $dt$,
the two processes do not interfere with each other (up to order $dt$)
, so that
\begin{multline}
\rho_{s}(t+dt)\\
=\left\langle M_{0}U_{t}\rho_{s}(t)U_{t}^{\dagger}M_{0}^{\dagger}+M_{1}U_{t}\rho_{s}(t)U_{t}^{\dagger}M_{1}^{\dagger}\right\rangle _{\bm{\xi}(t)}+O(dt^{2})\\
=\rho_{s}(t)-i\omega\left[\sigma_{x},\rho_{s}(t)\right]dt\\
-\gamma_{ij}\left(\frac{1}{2}\left\{ \sigma_{i}\sigma_{j},\rho_{s}(t)\right\} -\sigma_{i}\rho_{s}(t)\sigma_{j}\right)dt\\
-\alpha\left(\frac{1}{2}\left\{ P_{1},\rho_{s}(t)\right\} -P_{1}\rho_{s}(t)P_{1}\right)dt+O(dt^{2}).\label{eq:combine_two_evolns}
\end{multline}
This master equation is equivalent to the following equation for the
Bloch vector, cf.~Eq.~(\ref{eq:system_gen_state}),
\begin{equation}
\frac{d\bm{s}}{dt}=\mathcal{L}\bm{s},\quad\mathcal{L}=\mathcal{L}_{0}+\mathcal{L}_{\gamma},\label{eq:system_evoln_diff_eq}
\end{equation}
where the evolution superoperator $\mathcal{L}$ (Liouvillian) is
decomposed into the noiseless part
\begin{equation}
\mathcal{L}_{0}=\begin{pmatrix}-\frac{\alpha}{2} & 0 & 0\\
0 & -\frac{\alpha}{2} & -2\omega\\
0 & 2\omega & 0
\end{pmatrix}\label{eq:superoperator}
\end{equation}
and the noise contribution
\begin{equation}
\mathcal{L}_{\gamma}=\begin{pmatrix}-2(\gamma_{22}+\gamma_{33}) & 2\gamma_{12} & 2\gamma_{13}\\
2\gamma_{12} & -2(\gamma_{11}+\gamma_{33}) & 2\gamma_{23}\\
2\gamma_{13} & 2\gamma_{23} & -2(\gamma_{11}+\gamma_{22})
\end{pmatrix}.\label{eq:noisy_part_of_the_superoperator}
\end{equation}

Equation~(\ref{eq:system_evoln_diff_eq}) is formally solved by exponentiating
the Liouvillian:

\begin{equation}
\bm{s}(t)=e^{\mathcal{L}t}\bm{s}(0).\label{eq:system_evol_eq_soln}
\end{equation}

\section{QZE in the absence of noise \label{sec:in_the_absence_of_any_noise}}

Before dealing with the effect of noise on the QZE, we review the
features of the QZE in the absence of noise \citep{Gong2004,Koshino2005,Streed2006,Xiao2006,Ruskov2006,Breuer2007,Paz-Silva2012,Layden2015,Zhang2019}.
These known results will serve as a benchmark to assess the effects
of noise on the QZE.\textcolor{orange}{{} }A convenient quantifier for
this goal is the survival probability, i.e. the probability that the
system would be found in its initial state ($\ket 0$, cf.~Eq.~(\ref{eq:system_initial_state})),
when measured projectively at time $t$. This is given by

\begin{equation}
\mathcal{P}(t)=\textrm{Tr}[\rho_{s}(t)\rho_{s}(0)]=\frac{1+z(t)}{2}.\label{eq:survival_prob_def}
\end{equation}

In the absence of noise, $\gamma_{ij}=0$, the evolution superoperator
$\mathcal{L}$ in Eq.~(\ref{eq:system_evoln_diff_eq}) reduces to

\begin{equation}
\mathcal{L}^{(\bar{\mathrm{n}})}=\mathcal{L}_{0}=\begin{pmatrix}-\frac{\alpha}{2} & 0 & 0\\
0 & -\frac{\alpha}{2} & -2\omega\\
0 & 2\omega & 0
\end{pmatrix},\label{eq:superoperator_no_noise}
\end{equation}
where the superscript $(\bar{\mathrm{n}})$ represents the absence
of noise. The evolution of the Bloch vector $\bm{s}(t)$ is obtained
explicitly by diagonalizing the Liouvillian, $\mathcal{L}^{(\bar{\mathrm{n}})}=W^{(\bar{\mathrm{n}})}\Lambda^{(\bar{\mathrm{n}})}\left(W^{(\bar{\mathrm{n}})}\right)^{-1}$.
Here $\Lambda^{(\bar{\mathrm{n}})}$ is the diagonal matrix of eigenvalues
of $\mathcal{L}^{(\bar{\mathrm{n}})}$, given by

\begin{align}
\lambda_{1}^{(\bar{\mathrm{n}})}= & -\frac{\alpha}{2},\label{eq:lambda1_no_noise}\\
\lambda_{2}^{(\bar{\mathrm{n}})}= & -\frac{1}{4}(\alpha+\sqrt{\alpha^{2}-64\omega^{2}}),\label{eq:lambda2_no_noise}\\
\lambda_{3}^{(\bar{\mathrm{n}})}= & -\frac{1}{4}(\alpha-\sqrt{\alpha^{2}-64\omega^{2}}).\label{eq:lambda3_no_noise}
\end{align}
Then $\bm{s}(t)=W^{(\bar{\mathrm{n}})}\exp(\Lambda^{(\bar{\mathrm{n}})}t)\left(W^{(\bar{\mathrm{n}})}\right)^{-1}\bm{s}(0)$,
and, consequently, the survival probability reads
\begin{multline}
\mathcal{P}^{(\bar{\mathrm{n}})}(t)=\frac{1}{2}\left(1+e^{-\frac{\alpha t}{4}}\left(\cosh\frac{t\sqrt{\alpha^{2}-64\omega^{2}}}{4}\right.\right.\\
\left.+\left.\frac{\alpha}{\sqrt{\alpha^{2}-64\omega^{2}}}\sinh\frac{t\sqrt{\alpha^{2}-64\omega^{2}}}{4}\right)\right)\label{eq:survival_prob_no_noise}
\end{multline}
For any finite measurement strength $\alpha$, the probability asymptotically
decays to $1/2$ at $t\rightarrow\infty$. However, the rate of decay
depends on the measurement strength. In the limit of infinitely frequent
\emph{projective} measurements, $\alpha\rightarrow\infty$, the survival
probability is equal to 1 at all times, so the system never leaves
its initial state, which corresponds to the perfect QZE. For sufficiently
weak measurements, $\alpha<8\omega$, the survival probability exhibits
not only decay, but also oscillations.

In the following sections, we analyze the effect of the Hamiltonian
noise on the survival probability. We are specifically interested
in three aspects of its behavior:

(i) The short-time behavior. At $t\rightarrow0$,
\begin{equation}
\mathcal{P}^{(\bar{\mathrm{n}})}(t\rightarrow0)=1-\omega^{2}t^{2}+\frac{\alpha\omega^{2}t^{3}}{6}+O(t^{4}).\label{eq:surviavl_prob_short_time_behav_no_noise}
\end{equation}
We see that for any measurement strength, the probability initially
decays quadratically in time. The presence of measurements increases
the survival probability at short times, however, this is a cubic
order effect. Note that in this limit $\alpha t\ll1$ is implicitly
assumed, hence the freezing of the state ($\mathcal{P}^{(\bar{\mathrm{n}})}\equiv1$)
as $\alpha\to\infty$ is not apparent.

\begin{figure}
\begin{centering}
\includegraphics[width=1\columnwidth]{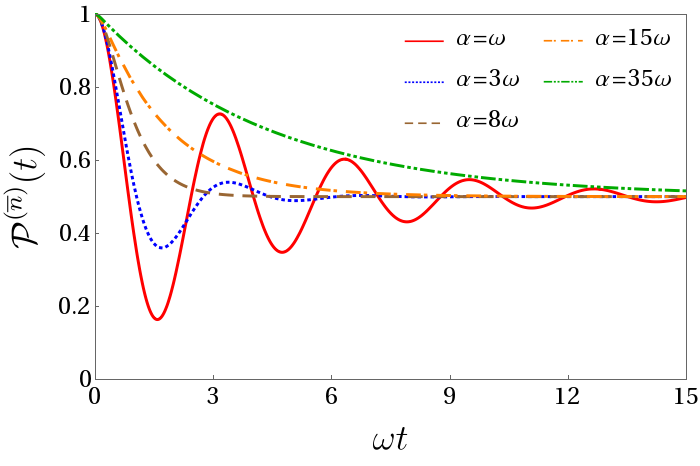}
\par\end{centering}
\raggedright{}\caption{\label{fig:survival_prob_no_noise}The behavior of the survival probability
$\mathcal{P}^{(\bar{\mathrm{n}})}(t)$ for different measurement strengths
in the absence of noise. The survival probability has an oscillatory
behavior for $\alpha<8\omega$ and non-oscillatory behavior for $\alpha\geqslant8\omega.$}
\end{figure}

(ii) The oscillations of the survival probability. The survival probability
oscillates as a function of time for $\alpha<8\omega$, which is a
consequence of the eigenvalues $\lambda_{2}^{(\bar{\mathrm{n}})}$
and $\lambda_{3}^{(\bar{\mathrm{n}})}$ being complex. The oscillations
vanish for $\alpha\geqslant8\omega$ (cf.~Fig.~\ref{fig:survival_prob_no_noise}).
The value of $\alpha=8\omega=\alpha_{\mathrm{exc}}^{(\bar{\mathrm{n}})}$,
where the Liouvillian's eigenvalues become degenerate, $\lambda_{2}^{(\bar{\mathrm{n}})}=\lambda_{3}^{(\bar{\mathrm{n}})}$,
is an exceptional point of the evolution superoperator \citep{Hatano2019,Minganti2019,Minganti2020}
and it can be identified as the critical measurement strength for
the onset of the QZE regime \citep{Li2014a,Snizhko2020d}.

\begin{figure}
\begin{centering}
\includegraphics[width=1\columnwidth]{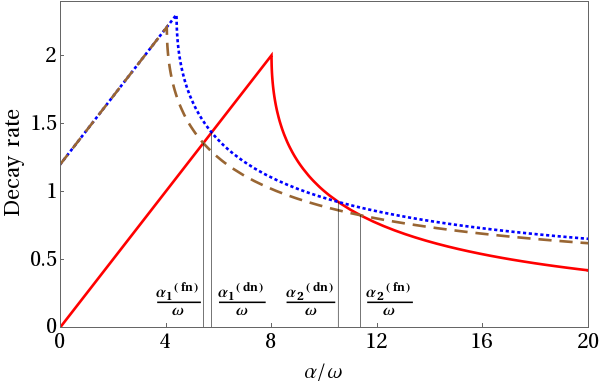}
\par\end{centering}
\caption{The behavior of the long-time decay rate of the survival probability
as a function of the measurement strength, $\alpha$, --- in the
absence of noise (solid red), when only the diagonal noise $\gamma_{ii}$
is present (dotted blue), and in the presence of both the diagonal
and off-diagonal noise (dashed brown). The decay rates are maximum
at respective exceptional points. It can be seen that, depending on
the measurement strength, noise can enhance or suppress the decay
rate. The presence of off-diagonal noise component increases the interval
over which the decay rate is suppressed compared to the case of diagonal
noise. The solid vertical lines mark the measurement strengths that
define the interval over which the decay rate is suppressed by the
noise, cf.~Eqs.~(\ref{eq:alphap_diag_noise}--\ref{eq:alphapp_diag_noise})
and (\ref{eq:alphap_full_noise}--\ref{eq:alphapp_full_noise}).
The plots are obtained for $\gamma_{11}=0.05\,\omega$, $\gamma_{22}=0.1\,\omega$,
$\gamma_{12}=\gamma_{13}=0$, $\gamma_{23}=0.3\,\omega$, and $\gamma_{33}=\omega$.}

\label{Fig:decay_rate_vs_alpha}
\end{figure}

(iii) The long-time behavior. At $t\rightarrow\infty$, the behavior
is determined by the slowest decaying eigenstate of the evolution
superoperator, which is the one associated with eigenvalue $\lambda_{3}^{(\bar{\mathrm{n}})}$.
Thus at long times, the survival probability can be written as
\begin{equation}
\mathcal{P}^{(\bar{\mathrm{n}})}(t\rightarrow\infty)=\frac{1}{2}+e^{-t|\mathrm{Re}\,\lambda_{3}^{(\bar{\mathrm{n}})}|}\times f(t),
\end{equation}
where $f(t)$ is either a constant or a bounded oscillating function.
Note that the decay rate $|\mathrm{Re}\,\lambda_{3}^{(\bar{\mathrm{n}})}$|,
for a fixed $\omega$, exhibits a maximum at $\alpha=8\omega=\alpha_{\mathrm{exc}}^{(\bar{\mathrm{n}})}$,
as shown in Fig~\ref{Fig:decay_rate_vs_alpha}. That is, the survival
probability long-time decay rate decreases with increasing $\alpha$
at $\alpha>8\omega$, while the behavior is opposite at $\alpha<8\omega$.

Therefore, the effect of the Hamiltonian noise on the QZE can be assessed
by three quantifiers: (i) the amount of suppression of the survival
probability at short times, (ii) the shift of the exceptional point,
and (iii) the effect on the survival probability's long-time decay
rate.

\section{Effect of noise on QZE\label{sec:the_effect_of_noise}}

We are now in a position to analyze the effect of Hamiltonian noise
on the QZE. This is generally described by six independent parameters
$\left\{ \gamma_{ij}\right\} $ with $i\leqslant j$, cf. Eq. (\ref{eq:noisy_part_of_the_superoperator}).
We start by considering the case of diagonal noise $\gamma_{ij}\propto\delta_{ij}$,
which demonstrates all the qualitative features appearing due to noise,
cf. Sec.~\ref{subsec:Diagonal_noise}. In Sec.~\ref{subsec:Diagonal_as_well_off_diagonal_noise},
we include the off-diagonal terms and discuss their effect.

\subsection{Diagonal noise \label{subsec:Diagonal_noise}}

The diagonal noise is defined by $\gamma_{ij}=\gamma_{ii}\delta_{ij}$
with $\gamma_{ii}\geqslant0$. Physically, this may originate from
the fluctuations of the energy difference between $\ket 0$ and $\ket 1$
states ($\gamma_{33}\neq0$), the fluctuations of the Rabi frequency
$\omega$ ($\gamma_{11}\neq0$), or the fluctuations of the Hamiltonian
direction in the $xy$ plane ($\gamma_{11}\neq0$ and $\gamma_{22}\neq0$).
The Liouvillian $\mathcal{L}$ (\ref{eq:system_evoln_diff_eq}) then
takes the form

\begin{multline}
\mathcal{L}^{(\mathrm{dn})}=\mathcal{L}_{0}+\mathcal{L}_{\gamma}\\
=\begin{pmatrix}-\frac{\alpha+4(\gamma_{22}+\gamma_{33})}{2} & 0 & 0\\
0 & -\frac{\alpha+4(\gamma_{11}+\gamma_{33})}{2} & -2\omega\\
0 & 2\omega & -2(\gamma_{11}+\gamma_{22})
\end{pmatrix}.\label{eq:superoperator_diagonal_noise}
\end{multline}
with the corresponding eigenvalues

\begin{eqnarray}
\lambda_{1}^{(\mathrm{dn})} & = & -\frac{\alpha_{\mathrm{dn}}}{2}-4\gamma_{22},\label{eq:lambda_1_diag_noise}\\
\lambda_{2}^{(\mathrm{dn})} & = & -\frac{1}{4}\biggl(\alpha_{\mathrm{dn}}+8(\gamma_{11}+\gamma_{22})+\sqrt{\alpha_{\mathrm{dn}}^{2}-64\omega^{2}}\biggr),\label{eq:lambda_2_diag_noise}\\
\lambda_{3}^{(\mathrm{dn})} & = & -\frac{1}{4}\biggl(\alpha_{\mathrm{dn}}+8(\gamma_{11}+\gamma_{22})-\sqrt{\alpha_{\mathrm{dn}}^{2}-64\omega^{2}}\biggr).\label{eq:lambda_3_diag_noise}
\end{eqnarray}
Here $\alpha_{\mathrm{dn}}=\alpha-4(\gamma_{22}-\gamma_{33})$ plays
the role of the renormalized measurement strength in some of the observables
(see below). With the system's initial state corresponding to $\ket 0$,
cf.~Eq.~(\ref{eq:system_initial_state}), one obtains the survival
probability

\begin{widetext}

\begin{equation}
\mathcal{P}^{(\mathrm{dn})}(t)=\frac{1}{2}\left(1+e^{-\frac{t(\alpha_{\mathrm{dn}}+8(\gamma_{11}+\gamma_{22}))}{4}}\left(\cosh\frac{t\sqrt{\alpha_{\mathrm{dn}}^{2}-64\omega^{2}}}{4}+\frac{\alpha_{\mathrm{dn}}}{\sqrt{\alpha_{\mathrm{dn}}^{2}-64\omega^{2}}}\sinh\frac{t\sqrt{\alpha_{\mathrm{dn}}^{2}-64\omega^{2}}}{4}\right)\right).\label{eq:survival_prob_with_diagonal_noise}
\end{equation}
\end{widetext}

In the limit $\alpha\to\infty$, $\mathcal{P}^{(\mathrm{dn})}(t)=\frac{1}{2}\left[1+e^{-2(\gamma_{11}+\gamma_{22})t}\right]$.
Thence, the noise will generically prevent a full freezing of the
state with $\mathcal{P}^{(\mathrm{dn})}(t)\to1$ in the ideal strong
measurement case, as was previously pointed out in Ref.~\citep{Gurvitz2003}.
However, for realistic finite values of the measurement strength,
the presence of noise can alter the survival probability behavior
and the critical measurement strength both favorably and not. First,
it is evident from Eq. (\ref{eq:superoperator_diagonal_noise}) that
$\gamma_{33}$ acts solely to renormalize the measurement strength
$\alpha$ (in agreement with the results of Refs.~\citep{Nakazato1999,Kofman1999}).
Therefore, the noise along the $z$ axis enhances the effective measurement
strength and thence the QZE. This has an intuitive explanation: A
non-zero $\sigma_{z}$ term in the Hamiltonian would reduce the size
of Rabi oscillations enabled by the $\omega\,\sigma_{x}$ term, thus
enhancing the survival probability; $\gamma_{33}\neq0$ corresponds
to having a fluctuating $\sigma_{z}$ term in the Hamiltonian that
similarly enhances the survival probability.\textcolor{orange}{{} }The
effect of $\gamma_{11}$ is also rather clear: it induces dephasing
between $\sigma_{x}$ eigenstates, hence inducing an exponential decay
of the $z$ component of the Bloch vector. This counteracts the onset
of the QZE, and in fact it prevents the full freezing of the state
for $\alpha\to\infty.$ The noise along the $y$ axis has the opposite
effect of $\gamma_{33}$ in renormalizing the measurement strength,
and acts analogously to $\gamma_{11}$ in affecting the exponential
decay, so to counteract overall the QZE.

\begin{figure}
\centering{}\includegraphics[width=1\columnwidth]{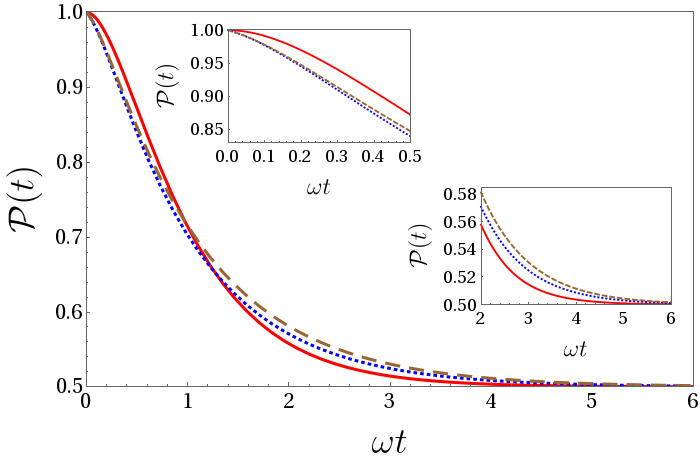}\caption{Time dependence of the survival probability $\mathcal{P}(t)$, cf.~Eq.~(\ref{eq:survival_prob_def}),
for the cases of no noise ($\gamma_{ij}=0$, solid red), diagonal
noise ($\gamma_{11}=0.05\,\omega$, $\gamma_{22}=0.1\,\omega$, $\gamma_{33}=\omega,$
dotted blue), and full noise (as diagonal noise plus $\gamma_{23}=0.3\,\omega$,
dashed brown). The insets focus on the short-time and long-time behavior.
The noise always decreases the survival probability at short times,
however, at long times it may actually increase it (the parameters
used in the plot correspond to the latter scenario). The plots are
obtained for $\alpha=8.5\,\omega$. The enhancement of the survival
probability at long times can be more significant; however, this only
happens for $\gamma_{33}$-dominated noise ($\gamma_{33}/\gamma_{11},\:\gamma_{33}/\gamma_{22}\gtrsim100$).\label{Fig:short_time_and_long_time_behavior_of_the_survival_prob}}
\end{figure}

Focusing specifically on the quantifiers of the QZE introduced in
Sec.~\ref{sec:in_the_absence_of_any_noise}, the short-time behavior
of the survival probability is given by
\begin{eqnarray}
\mathcal{P}^{(\mathrm{dn})}(t) & = & 1-(\gamma_{11}+\gamma_{22})t-\left(\omega^{2}-(\gamma_{11}+\gamma_{22})^{2}\right)t^{2}\nonumber \\
 & + & \frac{1}{6}\biggl(\omega^{2}\alpha-4(\gamma_{11}+\gamma_{22})^{3}\nonumber \\
 & + & 4\omega^{2}(3\gamma_{11}+2\gamma_{22}+\gamma_{33})\biggr)t^{3}+O(t^{4}).\label{eq:survival_prob_diag_noise_short_time}
\end{eqnarray}
Comparing this to the noiseless case in Eq.~(\ref{eq:surviavl_prob_short_time_behav_no_noise}),
we see that the presence of noise induces a linear-in-time decay.
Therefore, at short times, the QZE is suppressed by noise unless $\gamma_{11}=\gamma_{22}=0$,
cf.~Fig.~\ref{Fig:short_time_and_long_time_behavior_of_the_survival_prob}.
If this condition is satisfied, then the short-time behavior is equivalent
to that in the noiseless case with a renormalized measurement strength,
$\alpha\rightarrow\alpha+4\gamma_{33}>\alpha$. Therefore, noise along
the $z$ axis \emph{alone} does enhance the QZE at short times.

The analysis of the oscillatory dynamics of the survival probability,
controlled by the exceptional point of the Liouvillian spectrum, reveals
a rather different picture. The exceptional point (where $\lambda_{2}^{(\mathrm{dn})}=\lambda_{3}^{(\mathrm{dn})}$)
now occurs at

\begin{equation}
\alpha_{\mathrm{dn}}=8\omega\Longleftrightarrow\alpha=8\omega+4(\gamma_{22}-\gamma_{33})=\alpha_{\mathrm{exc}}^{(\mathrm{dn})}.\label{eq:exceptional_point_location_diag_noise}
\end{equation}
This implies that for a given $\omega$ the onset of the QZE regime
happens at a smaller measurement strength as long as $\gamma_{33}>\gamma_{22}$,
independently of $\gamma_{11}$. This condition for enhancing the
QZE is drastically different from the one found when considering the
short-time dynamics. 

Finally, consider the long-time decay rate of the survival probability.
The decay rate is given by $|\mathrm{Re}\,\lambda_{3}^{(\textrm{dn})}$|,
to be compared with the decay rate in the absence of noise, $|\mathrm{Re}\,\lambda_{3}^{(\bar{\mathrm{n}})}$|,
cf.~Eqs.~(\ref{eq:lambda_3_diag_noise}) and (\ref{eq:lambda3_no_noise})
respectively. We find that generically the noise increases the decay
rate, except when the following three conditions are satisfied simultaneously:
(a) \emph{$\gamma_{33}>\gamma_{22}$, }(b)\emph{ $\omega>(\gamma_{11}+\gamma_{22})(\gamma_{11}+\gamma_{33})/(\gamma_{33}-\gamma_{22})$,
}(c) the measurement strength, $\alpha$, falls into the interval

\begin{equation}
\alpha_{1}^{(\mathrm{dn})}<\alpha<\alpha_{2}^{(\mathrm{dn})},\label{eq:survival_prob_enhancement_cond_on_meas_coup_diag_noise}
\end{equation}
 with

\begin{equation}
\alpha_{1}^{(\mathrm{dn})}=4\biggl(\gamma_{22}-\gamma_{33}+\sqrt{4\omega^{2}+(2\gamma_{11}+\gamma_{22}+\gamma_{33})^{2}}\biggr),\label{eq:alphap_diag_noise}
\end{equation}
\begin{eqnarray}
\alpha_{2}^{(\mathrm{dn})} & = & 2\biggl(\gamma_{22}-\gamma_{33}\nonumber \\
 & + & (2\gamma_{11}+\gamma_{22}+\gamma_{33})\sqrt{1+\frac{4\omega^{2}}{(\gamma_{11}+\gamma_{22})(\gamma_{11}+\gamma_{33})}}\biggr).\nonumber \\
\label{eq:alphapp_diag_noise}
\end{eqnarray}
Such suppression of the decay rate (enhancement of the long-time survival
probability) is illustrated in Figs.~\ref{Fig:decay_rate_vs_alpha}
and \ref{Fig:short_time_and_long_time_behavior_of_the_survival_prob}.

We thus see that the conditions for enhancing the QZE by the noise
differ depending on the quantity under consideration. The short-time
survival probability is almost always suppressed by noise. The onset
of the QZE regime (i.e., the exceptional point) happens at a lower
measurement-strength when $\gamma_{33}>\gamma_{22}$. The enhancement
of the long-time survival probability agrees with that obtained from
the exceptional point shift in the requirement of $\gamma_{33}>\gamma_{22}$.
At the same time, the long-time survival probability enhancement poses
extra conditions: the noiseless Hamiltonian should be sufficiently
strong compared to some noise-related quantity (b) and the measurement
strength should belong to an interval close to the exceptional point
(c).

\subsection{Generic noise \label{subsec:Diagonal_as_well_off_diagonal_noise}}

We now analyze the case of generic noise with all $\gamma_{ij}\neq0$.
We focus on the regime of sufficiently weak noise, $\abs{\gamma_{ij}}\ll\alpha,\omega$.
When $\alpha\rightarrow\infty$, the $z$ component of the Bloch vector
is effectively decoupled from the $x$ and $y$ components, cf.~Eqs.~(\ref{eq:system_evoln_diff_eq}--\ref{eq:noisy_part_of_the_superoperator}),
implying that the effect of the off-diagonal noise components onto
the survival probability can be neglected. Therefore, we focus on
sufficiently small $\alpha$. For sufficiently small $\alpha$, one
can compute the Liouvillian eigenvalues perturbatively in $\gamma_{12}/\abs{\lambda_{1}^{(\bar{\mathrm{n}})}-\lambda_{2}^{(\bar{\mathrm{n}})}}\ll1$
and $\gamma_{13}/\abs{\lambda_{1}^{(\bar{\mathrm{n}})}-\lambda_{3}^{(\bar{\mathrm{n}})}}\ll1$
and show that these parameters only contribute at the second order
of perturbation theory. Thus, the effect of cross-correlations between
the noise along the $x$ and the other two axes is negligibly small
in the considered regime.

Diagonalizing the evolution superoperator (\ref{eq:system_evoln_diff_eq})
and neglecting the contributions of $\gamma_{12}$ and $\gamma_{13}$,
we obtain

\begin{eqnarray}
\lambda_{1}^{(\mathrm{fn})} & = & -\frac{\alpha_{\mathrm{dn}}}{2}-4\gamma_{22},\label{eq:lambda_1_full_noise}\\
\lambda_{2}^{(\mathrm{fn})} & = & -\frac{1}{4}\biggl(\alpha_{\mathrm{dn}}+8(\gamma_{11}+\gamma_{22})+\sqrt{\alpha_{\mathrm{dn}}^{2}-64\left(\omega^{2}-\gamma_{23}^{2}\right)}\biggr),\nonumber \\
\label{eq:lambda_2_full_noise}\\
\lambda_{3}^{(\mathrm{fn})} & = & -\frac{1}{4}\biggl(\alpha_{\mathrm{dn}}+8(\gamma_{11}+\gamma_{22})-\sqrt{\alpha_{\mathrm{dn}}^{2}-64\left(\omega^{2}-\gamma_{23}^{2}\right)}\biggr).\nonumber \\
\label{eq:lambda_3_full_noise}
\end{eqnarray}
Note that if $\gamma_{12}=\gamma_{13}=0$, our analysis is exact and
is valid for arbitrary strength noise.

The survival probability is then obtained as\begin{widetext}

\begin{multline}
\mathcal{P}^{(\mathrm{fn})}=\frac{1}{2}\biggl(1+e^{-\frac{t(\alpha_{\mathrm{dn}}+8(\gamma_{11}+\gamma_{22}))}{4}}\biggl(\cosh\frac{t\sqrt{\alpha_{\mathrm{dn}}^{2}-64\left(\omega^{2}-\gamma_{23}^{2}\right)}}{4}\\
+\frac{\alpha_{\mathrm{dn}}}{\sqrt{\alpha_{\mathrm{dn}}^{2}-64\left(\omega^{2}-\gamma_{23}^{2}\right)}}\sinh\frac{t\sqrt{\alpha_{\mathrm{dn}}^{2}-64\left(\omega^{2}-\gamma_{23}^{2}\right)}}{4}\biggr)\biggr).\label{eq:survival_prob_full_noise}
\end{multline}

\end{widetext} The short-time behavior of the survival probability
is thus the same as in the case of diagonal noise (\ref{eq:survival_prob_diag_noise_short_time}),
modulo replacing $\omega^{2}\rightarrow\omega^{2}-\gamma_{23}^{2}$,
\begin{eqnarray}
\mathcal{P}^{(\mathrm{fn})} & = & 1-(\gamma_{11}+\gamma_{22})t-\left(\omega^{2}-\gamma_{23}^{2}-(\gamma_{11}+\gamma_{22})^{2}\right)t^{2}\nonumber \\
 & + & \frac{1}{6}\biggl((\omega^{2}-\gamma_{23}^{2})\left(\alpha+4(3\gamma_{11}+2\gamma_{22}+\gamma_{33})\right)\nonumber \\
 & - & 4(\gamma_{11}+\gamma_{22})^{3}\biggr)t^{3}+O(t^{4}).\label{eq:survival_prob_full_noise_small_time}
\end{eqnarray}
Hence, the short-time survival probability is always reduced by the
noise unless $\gamma_{11}=\gamma_{22}=0$ (note that the requirement
of $\gamma_{ij}$ being a positive semidefinite matrix implies that
in this case $\gamma_{ij}=0$ unless $i=j=3$). Note also the subleading
$t^{2}$ term, where the presence of $\gamma_{23}$ enhances the survival
probability, cf.~Fig.~\ref{Fig:short_time_and_long_time_behavior_of_the_survival_prob}.

The exceptional point happens when

\begin{equation}
\alpha=4\left(2\sqrt{\omega^{2}-\gamma_{23}^{2}}+\gamma_{22}-\gamma_{33}\right)=\alpha_{\mathrm{exc}}^{(\mathrm{fn})}.
\end{equation}
This is a smaller value of $\alpha$ than $\alpha_{\mathrm{exc}}^{(\bar{\mathrm{n}})}=8\omega$
provided that $\gamma_{33}>\gamma_{22}$, or if $\gamma_{33}<\gamma_{22}$
and $4\omega<(\gamma_{22}-\gamma_{33})+4\gamma_{23}^{2}/(\gamma_{22}-\gamma_{33})$.
When $\abs{\gamma_{ij}}\ll\omega$, the last condition on $\omega$
can only be satisfied when $\gamma_{22}-\gamma_{33}\ll\abs{\gamma_{23}}$. 

Finally, for the effect of noise in the long-time limit, we compare
$|\mathrm{Re}\,\lambda_{3}^{(\bar{\mathrm{n}})}$| and $|\mathrm{Re}\,\lambda_{3}^{(\mathrm{fn})}|$,
cf.~Eqs.~(\ref{eq:lambda3_no_noise}) and (\ref{eq:lambda_3_full_noise}).
Similarly to the case of diagonal noise, the long-time decay rate
is reduced if and only if the following three conditions are satisfied:
(a) $\gamma_{33}>\gamma_{22}$, (b) $\omega>((\gamma_{11}+\gamma_{22})(\gamma_{11}+\gamma_{33})-\gamma_{23}^{2})/(\gamma_{33}-\gamma_{22})$,
(c) the measurement strength belongs to the interval

\begin{equation}
\alpha_{1}^{(\mathrm{fn})}<\alpha<\alpha_{2}^{(\mathrm{fn})},\label{eq:survival_prob_enhancement_cond_on_meas_coup_full_noise}
\end{equation}
 with \begin{widetext}

\begin{eqnarray}
\alpha_{1}^{(\mathrm{fn})} & = & 4\biggl(\gamma_{22}-\gamma_{33}+\sqrt{4(\omega^{2}-\gamma_{23}^{2})+(2\gamma_{11}+\gamma_{22}+\gamma_{33})^{2}}\biggr),\label{eq:alphap_full_noise}\\
\alpha_{2}^{(\mathrm{fn})} & = & 2\biggl((\gamma_{22}-\gamma_{33})\left(1-\frac{\gamma_{23}^{2}}{(\gamma_{11}+\gamma_{22})(\gamma_{11}+\gamma_{33})}\right)\nonumber \\
 & + & (2\gamma_{11}+\gamma_{22}+\gamma_{33})\sqrt{\left(1-\frac{\gamma_{23}^{2}}{(\gamma_{11}+\gamma_{22})(\gamma_{11}+\gamma_{33})}\right)^{2}+\frac{4\omega^{2}}{(\gamma_{11}+\gamma_{22})(\gamma_{11}+\gamma_{33})}}\biggr).\label{eq:alphapp_full_noise}
\end{eqnarray}
\end{widetext}Note that the off-diagonal noise enhances the QZE under
a wider range of conditions. Indeed, when $\gamma_{23}\neq0$, the
restriction on $\omega$ is weaker than for the diagonal noise with
the same $\gamma_{ii}$. Further, for $\omega>(\gamma_{11}+\gamma_{22})(\gamma_{11}+\gamma_{33})/(\gamma_{33}-\gamma_{22})$
at which the diagonal noise allows for reducing the decay rate, adding
the off-diagonal component increases the relevant interval of measurement
strengths: $\alpha_{1}^{(\mathrm{fn})}<\alpha_{1}^{(\mathrm{dn})}$
and $\alpha_{2}^{(\mathrm{fn})}>\alpha_{2}^{(\mathrm{dn})}$, as illustrated
in Fig.~\ref{Fig:decay_rate_vs_alpha}.

We thus see that the off-diagonal noise components do not qualitatively
change the effect of noise on the QZE. It is interesting to note,
though, that correlations between the $y$ and the $z$ components
of noise ($\gamma_{23}\neq0$) tend to enhance the QZE compared to
the purely diagonal noise.

\section{conclusion \label{sec:conclusion}}

In this work we have investigated the effect of noise in the system
Hamiltonian on the QZE behavior induced by continuous partial measurement.
We have found a quite rich behavior: the effect is significantly different
when looking at different QZE quantifiers. Specifically, we have investigated
the effect of noise on the short-time and long-time survival probabilities
and on the critical measurement strength determining the onset of
the QZE regime. We have found that the short-time survival probability
is essentially always reduced by noise, except for very fine tuned
conditions, namely, for a fluctuating term that commutes with the
measured observable. The onset of the QZE can be shifted towards larger
or smaller measurement strengths depending on the details of the noise
and the averaged Hamiltonian parameters. Finally, the long-time survival
probability can also be increased or decreased, yet this is not determined
by the properties of the noise only. The same noise can enhance or
suppress the long-time survival probability depending on the measurement
strength and the noiseless part of the system's Hamiltonian. Notably,
the conditions for a shift of the QZE onset towards a lower measurement
strength differ from those required to enhance the long time survival
probability. Our results can be relevant for QZE-based protocols in
systems subject to fluctuations, e.g., to optimize working points
in parameter space in order to enhance the desired features of the
QZE.

In this work we focused on Hermitian Hamiltonians. Interplay of non-Hermitian
Hamiltonian noise \citep{Burgarth2017} with measurement dynamics
may be of interest for applications in non-Hermitian systems.
\begin{acknowledgments}
P.K. and K.S. acknowledge funding by the Deutsche Forschungsgemeinschaft
(DFG, German Research Foundation) -- Projektnummer 277101999 --
TRR 183 (project C01) and Projektnummer EG 96/13-1, and by the Israel
Science Foundation (ISF). A.R. acknowledges the EPSRC via Grant No.
EP/P010180/1.
\end{acknowledgments}

\bibliography{bibliography}

\end{document}